\RequirePackage{fix-cm}
\documentclass[smallextended]{svjour3}       
\usepackage{amsmath}
\usepackage{amssymb}

\smartqed  
\usepackage{graphicx}
\begin{document}

\title{Discrete Phase Space, Relativistic Quantum Electrodynamics, and a Non-Singular Coulomb Potential
}

\author{Anadijiban Das         \and
        Rupak Chatterjee     \and
        Ting Yu
}


\institute{Anadijiban Das \at
              Department of Mathematics, Simon Fraser University, Burnaby, British Columbia, V5A 1S6, Canada \\
              \email{Das@sfu.ca}           
           \and
           Rupak Chatterjee \at
              Center for Quantum Science and Engineering\\
Department of Physics, Stevens Institute of Technology, Castle Point on the Hudson, Hoboken, NJ 07030, USA \\ \email{Rupak.Chatterjee@Stevens.edu}  
		\and
           Ting Yu \at
              Center for Quantum Science and Engineering\\
Department of Physics, Stevens Institute of Technology, Castle Point on the Hudson, Hoboken, NJ 07030, USA \\ \email{Ting.Yu@Stevens.edu}  
}

\date{Received: date / Accepted: date}

\maketitle

\begin{abstract}
This paper deals with \textit{the relativistic, quantized electromagnetic and Dirac field equations} in the arena of discrete phase space and continuous time. The mathematical formulation involves \textit{partial difference equations}. In the consequent relativistic quantum electrodynamics, the corresponding Feynman diagrams and 
$S^{\#}$-matrix elements are derived. In the special case of electron-electron scattering (M\o ller scattering), the explicit second order element $\langle f | S^{\#}_{(2)} |i \rangle$ is deduced. Moreover, assuming the slow motions for two external electrons, the approximation of $\langle f | S^{\#}_{(2)} |i \rangle$ yields a \textit{divergence-free Coulomb potential.}

\end{abstract}

\section{Introduction}
Partial difference equations have been studied \cite{Garabedian} for many years to investigate problems of mathematical physics \cite{Hamilton,Schiff,DuffinI,DuffinII,DasI}. Quantum mechanics has been exactly represented in phase space continuum with the usual time variable \cite{Wigner}. In recent years, an exact representation of quantum mechanics has been introduced in the discrete phase space and continuous time arena \cite{DasII,DasIII,DasIV,Gorski}. This representation involves a characteristic length $ l > 0$. Contrary to the usual expectations, this representation is exactly relativistic.

Furthermore, this representation can be elevated to the second quantization of free fields, interacting fields, and the new $S^{\#}$-matrix theory \cite{DasV,DasVI,DasVII}.

In this paper, we discuss a new discrete phase space-continuous time formulation of quantum electrodynamics, the $S^{\#}$-matrix, and the corresponding Feynman prescriptions \cite{DasVII}. We specifically concentrate on  second order electron-electron scattering (or M\o ller scattering). Furthermore, in the low momenta approximation of two external electrons, we derive a new Coulomb potential which is devoid of any singularity whatsoever. We hope that a future precise experiment can verify the validity of this new Coulomb potential.
 
\section{Notations and preliminary definitions}

There exists a characteristic length $l >0$ in this theory. We choose physical units such that $\hbar = c= l =1$ and express all physical quantities as dimensionless numbers. Greek indices take values from $\{1,2,3,4\}$ whereas the Roman indices take values from $\{1,2,3\}$. Einstein's summation convention is adopted in both cases. We denote the flat space-time metric by $\eta_{\mu \nu}$ with the corresponding diagonal matrix $[\eta_{\mu \nu}] := diag[1,1,1,-1]$. Therefore, we use a signature of +2 in this paper. An element of our discrete space phase and continuous time is expressed as $(\mathbf{n},x^4) \equiv (n^1, n^2, n^3, t) \in \mathbb N^3 \times \mathbb R, n^j \in \mathbb N$ for $j \in {1,2,3}$ and $x^4 \equiv t \in \mathbb R$.

Let a function $f$ from $\mathbb N^3 \times \mathbb R$ into $\mathbb R$ or $\mathbb C$ be denoted as $f(\mathbf{n}, t) = f(n^1, n^2, n^3, t)$. The various \textit{partial difference operators and the partial differential operators} are shown below \cite{DasV,DasVI}:  

\begin{subequations}
\begin{align}
\Delta_j f(\mathbf{n}, t) & := f(...,n^j+1, ..., t) - f(...,n^j, ..., t)   \\
\Delta_j^{'} f(\mathbf{n}, t) & := f(...,n^j, ..., t) - f(...,n^j-1, ..., t)   \\
\Delta_j^{\#} f(\mathbf{n}, t) & := \dfrac{1}{\sqrt{2}} \left[
 \sqrt{n^j+1} f(...,n^j+1, ..., t) 
 - \sqrt{n^j} f(...,n^j-1, ..., t)  \right] \\
\partial_t f(\mathbf{n}, t) & := \dfrac{\partial}{\partial t} [f(\mathbf{n}, t)]
\end{align}
\end{subequations}

We denote Hermite polynomials and some useful properties by the following \cite{DasV,DasVI}:
\begin{subequations}
\begin{align}
& H_{n^j}(k_j) := (-1)^{n^j} e^{(k_j)^2}  \dfrac{d^{n^j}}{(dk_j)^{n^j}}[e^{-(k_j)^2}]   \\
& \xi_{n^j}(k_j)  := \dfrac{(i)^{n^j} e^{-(k_j)^2/2} H_{n^j}(k_j)}{(\pi)^{1/4}2^{(n^j/2)}\sqrt{(n^j)!}} \\ 
& \int_{{\mathbb R}^3} \left\{ \prod_{j=1}^3 \left[ \xi_{n^j}(k_j)  \overline{\xi_{\hat{n}^j}(k_j)}\right] \right\}   dk_1 dk_2 dk_3  
= \delta_{n^1 \hat{n}^1} \delta_{n^2 \hat{n}^2} \delta_{n^3 \hat{n}^3}  =: \delta^3_{\mathbf{n} \hat{\mathbf{n}}} \\
& -i \Delta^{\#}  \xi_{n^j}(k_j) = k_j \xi_{n^j}(k_j) 
\end{align}
\end{subequations}

\section{Quantization of the free relativistic electromagnetic wave field}

The electromagnetic four-potential wave field is denoted by $A^{\mu}(\mathbf{n},t)$. It is assumed to be operator-valued and satisfies the \textit{partial difference-differential equations} \cite{DasV,DasVI}: 
\begin{equation}
\delta^{ab} \Delta_a^{\#}\Delta_b^{ \#} A_{\sigma}(\mathbf{n},t) -(\partial_t)^2 A_{\sigma}(\mathbf{n},t) = 0
\end{equation}
We assume that the four-potential function $A_{\sigma}(\mathbf{n},t)$ also satisfies the Lorenz-gauge condition,
\begin{equation}
\Delta_b^{ \#} A^{b}(\mathbf{n},t) +\partial_t A^{4}(\mathbf{n},t) = 0
\end{equation}

For a classical electromagnetic field, the equations (3) and (4) with the following definitions 
$F_{ab}(\mathbf{n},t):= \Delta_a^{ \#}A_{b}(\mathbf{n},t)  - \Delta_b^{ \#}A_{a}(\mathbf{n},t)$, and
$F_{a4}(\mathbf{n},t):= \Delta_a^{ \#}A_{4}(\mathbf{n},t)  - \partial_t A_{a}(\mathbf{n},t)$ yield exactly the electromagnetic field equations in the discrete phase space-continuous time arena. However, after the second quantization, the operator version of (4) poses a problem. One possible solution is to replace (4) with a weaker expectation value equation:
\begin{equation}
\langle \Psi_{(p)} | \Delta_b^{ \#} A^{b}(\mathbf{n},t) +\partial_t A^{4}(\mathbf{n},t)| \Psi_{(p)} \rangle  = 0
\end{equation}
Here, $\Psi_{(p)}$ indicates a physically admissible Hilbert space vector.

The momentum-energy four-vector of an external photon belongs to a four-dimensional Minkowskian vector space. Therefore, we can denote this entity by 
\begin{equation}
(k_1, k_2, k_3, k_4) :=(\mathbf{k}, k_4),
\end{equation}
where
\begin{equation}
\eta^{\mu \nu} k_\mu k_\nu =0, (k_4 )^2 =\delta^{ab} k_a k_b ,
\end{equation}
and
\begin{equation}
\nu = \nu(\mathbf{k}) := \sqrt{\delta^{ab} k_a k_b } = + \sqrt{\mathbf{k} \cdot \mathbf{k}} > 0.
\end{equation}
The symbol $\nu = \nu(\mathbf{k})$ physically stands for the frequency of the electromagnetic wave propagation.

We introduce four M-orthonormal or tetrad vectors by
\begin{subequations}
\begin{align}
e_{(\lambda)}^{\mu} (\mathbf{k}) & := \tilde{e}_{(\lambda)}^{\mu} (\mathbf{k}, \nu(\mathbf{k})) ,\\
\eta^{(\lambda \sigma)} e_{(\lambda)}^{\mu}(\mathbf{k}) e_{(\sigma)}^{\nu}(\mathbf{k}) & := \eta^{\mu \nu}, \\ 
\eta^{\mu \nu} e_{\mu}^{(\lambda)}(\mathbf{k}) e_{\nu}^{(\sigma)}(\mathbf{k}) & := \eta^{(\lambda \sigma)}.
\end{align}
\end{subequations}
We choose the spatial direction of the photon momentum propagator vector $\mathbf{k}$ along the third axis. A compatible selection of polarization vectors $e_{\mu}^{(\lambda)}(\mathbf{k})$ are furnished by \cite{Muirhead}
\begin{subequations}
\begin{align}
(k_1, k_2, k_3, k_4) & = (0, 0, k_3, \nu(\mathbf{k})) ,\\
e_{\mu}^{(\lambda)}(\mathbf{k}) & = \delta^{(\lambda)}_{\mu}, \\ 
k^{\mu} e_{\mu}^{(\lambda)}(\mathbf{k}) & = 0 \,\,\, \textbf{for} \,\,\, \lambda=1,2,\\
k^{\mu} e_{\mu}^{(3)}(\mathbf{k}) & = k^3, \,\,\, k^{\mu} e_{\mu}^{(4)}(\mathbf{k})  = -\nu(\mathbf{k})
\end{align}
\end{subequations}
The quantization of the electromagnetic four-potential involves Hermitian operators $A_{\sigma}(\mathbf{n},t)$ satisfying the partial differential-difference equations (3). A class of exact solutions of (3) is furnished by the integrals \cite{DasV,DasVI}:
\begin{subequations}
\begin{align}
& A_{\mu}(\mathbf{n},t) = A_{\mu}^{\dagger}(\mathbf{n},t)  = \int\limits_{{\mathbb R}^3} d^3\mathbf{k}\,\,\,[2\nu(\mathbf{k})]^{-1/2}  \notag\\
&\left\{a_{\mu}(\mathbf{k}) \left[\prod_{j=1}^{3} \xi_{n^j}(k_j) \right]  e^{-i\nu t}  + a^{\dagger}_{\mu}(\mathbf{k}) \left[\prod_{j=1}^{3} \overline{\xi_{n^j}(k_j)} \right] e^{i\nu t} \right\}  \\ 
& A_{\mu}^{(-)}(\mathbf{n},t)  = \int\limits_{{\mathbb R}^3} d^3\mathbf{k}\,\,\,[2\nu(\mathbf{k})]^{-1/2} \left\{a_{\mu}(\mathbf{k}) \left[\prod_{j=1}^{3} \xi_{n^j}(k_j) \right]  e^{-i\nu t} \right\} = \notag\\
&  \sum_{\lambda=1}^{2}\int\limits_{{\mathbb R}^3} d^3\mathbf{k}\,\,\,[2\nu(\mathbf{k})]^{-1/2} \left\{a_{(\lambda)}(\mathbf{k}) e_{\mu}^{(\lambda)}(\mathbf{k})\left[\prod_{j=1}^{3} \xi_{n^j}(k_j) \right]  e^{-i\nu t} \right\}  \\
& A_{\mu}^{(+)}(\mathbf{n},t)  = \int\limits_{{\mathbb R}^3} d^3\mathbf{k}\,\,\,[2\nu(\mathbf{k})]^{-1/2} \left\{a^{\dagger}_{\mu}(\mathbf{k}) \left[\prod_{j=1}^{3} \overline{\xi_{n^j}(k_j)} \right]  e^{+i\nu t} \right\} = \notag\\
&  \sum_{\lambda=1}^{2}\int\limits_{{\mathbb R}^3} d^3\mathbf{k}\,\,\,[2\nu(\mathbf{k})]^{-1/2} \left\{a^{\dagger}_{(\lambda)}(\mathbf{k}) e_{\mu}^{(\lambda)}(\mathbf{k})\left[\prod_{j=1}^{3} \overline{\xi_{n^j}(k_j)} \right]  e^{+i\nu t} \right\} \\
& A_{\mu}(\mathbf{n},t) =  A_{\mu}^{(-)}(\mathbf{n},t) + A_{\mu}^{(+)}(\mathbf{n},t) 
\end{align}
\end{subequations}
The operators $A_{\mu}^{(-)}(\mathbf{n},t)$ represent external photons terminating at $(\mathbf{n},t)$, whereas 
$A_{\mu}^{(+)}(\mathbf{n},t)$ represent external photons emanating from $(\mathbf{n},t)$.

The canonical quantization rules for operators $a_{\mu}(\mathbf{k})$ and $a^{\dagger}_{\mu}(\mathbf{k})$ are assumed to be the commutators:
\begin{subequations}
\begin{align}
[a_{\mu}(\mathbf{k}), a_{\nu}(\hat{\mathbf{k}})] & = [a^{\dagger}_{\mu}(\mathbf{k}), a^{\dagger}_{\nu}(\hat{\mathbf{k}})] =0\\
[a_{\mu}(\mathbf{k}), a^{\dagger}_{\nu}(\hat{\mathbf{k}})] & = -[a^{\dagger}_{\nu}(\hat{\mathbf{k}}), a_{\mu}(\mathbf{k})]= \eta_{\mu \nu} \delta^3 (\mathbf{k}-\hat{\mathbf{k}}) \mathbf{I}
\end{align}
\end{subequations}

Using (9) and (10), we obtain from (12)
\begin{subequations}
\begin{align}
[a^{(\lambda)}(\mathbf{k}), a^{(\hat{\lambda})}(\hat{\mathbf{k}})] & = [a^{\dagger (\lambda)}(\mathbf{k}), a^{\dagger (\hat{\lambda})}(\hat{\mathbf{k}})] =0\\
[a^{(\lambda)}(\mathbf{k}), a^{\dagger (\hat{\lambda})}(\hat{\mathbf{k}})] & = -[a^{\dagger (\hat{\lambda})}(\hat{\mathbf{k}}), a^{(\lambda)}(\mathbf{k})]= \eta^{(\lambda \hat{\lambda}) } \delta^3 (\mathbf{k}-\hat{\mathbf{k}}) \mathbf{I} \\
[a^{(\lambda)}(\mathbf{k}), a^{\dagger (\hat{\lambda})}(\hat{\mathbf{k}})] & = \delta^{(\lambda \hat{\lambda}) } \delta^3 (\mathbf{k}-\hat{\mathbf{k}}) \mathbf{I} \,\,\,\textit{for} \,\,\, \lambda, \hat{\lambda}  \in \{1,2\}.
\end{align}
\end{subequations}
In the sequel, we shall denote polarization indices only as $\lambda, \hat{\lambda}  \in \{1,2\}$. We use expressions (11) and commutation relations (12) to obtain
\begin{subequations}
\begin{align}
[A_{\mu}^{(+)}(\mathbf{n},t), A_{\nu}^{(+)}(\hat{\mathbf{n}},\hat{t})] & = [A_{\mu}^{(-)}(\mathbf{n},t), A_{\nu}^{(-)}(\hat{\mathbf{n}},\hat{t})] =0\\
[A_{\mu}^{(-)}(\mathbf{n},t), A_{\nu}^{(+)}(\hat{\mathbf{n}},\hat{t})] & = -i \eta_{\mu \nu } D_{(+)}(\mathbf{n},t;\hat{\mathbf{n}},\hat{t}) \mathbf{I} \\
[A_{\mu}^{(+)}(\mathbf{n},t), A_{\nu}^{(-)}(\hat{\mathbf{n}},\hat{t})] & = -i \eta_{\mu \nu } D_{(-)}(\mathbf{n},t;\hat{\mathbf{n}},\hat{t}) \mathbf{I} \\
[A_{\mu}(\mathbf{n},t), A_{\nu}(\hat{\mathbf{n}},\hat{t})] & = -i \eta_{\mu \nu } D(\mathbf{n},t;\hat{\mathbf{n}},\hat{t}) \mathbf{I} \\
[A_{\mu}(\mathbf{n},t), A_{\nu}(\hat{\mathbf{n}},\hat{t})]|_{t=\hat{t}} & = 0 \,\,\,\textit{for} \,\,\, \mathbf{n} \neq \hat{\mathbf{n}}
\end{align}
\end{subequations}
Here, $D_{(\pm)}(\mathbf{n},t;\hat{\mathbf{n}},\hat{t})$ and $D(\mathbf{n},t;\hat{\mathbf{n}},\hat{t})$ are non-singular Green's functions (or photon propagators) to be discussed in the Appendix. The equation (14e) indicates physically micro-causality in regards to measurements of two photons situated at two distinct discrete points of phase space at the same time.

\section{Quantization of the free relativistic electron-positron wave field}

Let us introduce one possible representation of $4 \times 4$ Dirac matrices with real and complex entries \cite{DasVIII} in the following way,  

\begin{equation}
\begin{array}{l}
 \gamma^{1} = [\gamma^{1}_{AB}] :=
\begin{bmatrix}
  0 & 0 & 0 & 1 \\
  0 & 0 & 1 & 0 \\
  0 & 1 & 0 & 0 \\
  1 & 0 & 0 & 0 \\
\end{bmatrix},
\gamma^{2} = [\gamma^{2}_{AB}] :=
\begin{bmatrix}
  0 & 0 & 0 & -i \\
  0 & 0 & i & 0 \\
  0 & -i & 0 & 0 \\
  i & 0 & 0 & 0 \\
\end{bmatrix} \\
 \gamma^{3} = [\gamma^{3}_{AB}] :=
\begin{bmatrix}
  0 & 0 & 1 & 0 \\
  0 & 0 & 0 & -1 \\
  1 & 0 & 0 & 0 \\
  0 & -1 & 0 & 0 \\
\end{bmatrix},
\gamma^{4} = [\gamma^{4}_{AB}] := 
\begin{bmatrix}
  -i & 0 & 0 & 0 \\
  0 & -i & 0 & 0 \\
  0 & 0 & i & 0 \\
  0 & 0 & 0 & i \\
\end{bmatrix}
\end{array}
\end{equation}
Here, the bispinor indices $A,B$ take values from $\{1,2,3,4\}$. The important properties of Dirac matrices are listed below:
\begin{subequations}
\begin{align}
\gamma^{a\dagger} = \gamma^{a} & , \gamma^{4\dagger} = -\gamma^{4}\\
\gamma^{\mu} \gamma^{\nu} +  \gamma^{\nu} \gamma^{\mu} & = 2 \eta^{\mu \nu} I_{(4 \times 4)}
\end{align}
\end{subequations}
The classical Dirac bispinor wave field $\psi(\mathbf{n},t)=[\psi^{A}(\mathbf{n},t)]$ is a $4 \times 1$ column vector field with entries of complex numbers. It physically represents a relativistic, classical electron-positron wave field defined over discrete phase space and continuous time. We introduce a corresponding conjugate   $1 \times 4$ row wave field as follows,
\begin{equation}
\tilde{\psi}(\mathbf{n},t) := i \psi^{\dagger}(\mathbf{n},t) \gamma^{4}
\end{equation}

The discrete phase space continuous time Dirac wave equations are furnished by
\begin{subequations}
\begin{align}
\gamma^a \Delta^{\#}_{a} \psi(\mathbf{n},t) + \gamma^{4} \partial_t \psi(\mathbf{n},t)  + m\psi(\mathbf{n},t) & = 0_{(4 \times 1)} \\
[\Delta^{\#}_{a} \tilde{\psi}(\mathbf{n},t)]\gamma^{a} + [\partial_t \tilde{\psi}(\mathbf{n},t)] \gamma^{4} -m\tilde{\psi}(\mathbf{n},t) & = 0_{(1 \times 4)}
\end{align}
\end{subequations}
Here, the positive constant $m>0$ represents the mass of an electron or a positron.

We now explore plane wave solutions of (18a) by the trial solution:
\begin{equation}
\psi(\mathbf{n},t) = \zeta(\mathbf{p}, p_4) \prod_{j=1}^{3} \xi_{n^j}(p_j)e^{ip_4 t}
\end{equation}
By substituting (19) into (18a), we derive that
\begin{subequations}
\begin{align}
& \eta^{\mu \nu}p_{\mu}p_{\nu} +m^2  = 0,\\
p_4=\pm E(\mathbf{p}), &\,\,\, E(\mathbf{p})=+\sqrt{\delta^{ab} p_a p_b +m^2} > 0
\end{align}
\end{subequations}
Using (20b), the four linearly independent solutions of (19) are given by
\begin{subequations}
\begin{align}
u_{(1)}(\mathbf{p}) = [(m+E)/2E]^{1/2}
\begin{bmatrix}
  1 \\
  0 \\
  -i(m+E)^{-1}p_3  \\
  -i(m+E)^{-1}(p_1 +ip_2)    \\
\end{bmatrix},\\
u_{(2)}(\mathbf{p}) = [(m+E)/2E]^{1/2}
\begin{bmatrix}
  0 \\
  1 \\
  -i(m+E)^{-1}(p_1 -ip_2)  \\
  i(m+E)^{-1}p_3    \\
\end{bmatrix},\\
v_{(1)}(\mathbf{p}) = [(m+E)/2E]^{1/2}
\begin{bmatrix}
  i(m+E)^{-1}p_3  \\
  -i(m+E)^{-1}(p_1 +ip_2)    \\
  1  \\
  0   \\
\end{bmatrix},\\
v_{(2)}(\mathbf{p}) = [(m+E)/2E]^{1/2}
\begin{bmatrix}
  i(m+E)^{-1}(p_1 -ip_2)  \\
  -i(m+E)^{-1}p_3    \\
  0  \\
  1   \\
\end{bmatrix}.
\end{align}
\end{subequations}
The above solutions (21) satisfy
\begin{subequations}
\begin{align}
& \tilde{u}_{(r)}(\mathbf{p}) \cdot u_{(s)}(\mathbf{p})  := \delta_{AB} \tilde{u}^{A}_{(r)}(\mathbf{p}) u^{B}_{(s)}(\mathbf{p}) 
 = - \tilde{v}_{(r)}(\mathbf{p}) \cdot v_{(s)}(\mathbf{p}) =\delta_{(rs)} \\
& \tilde{u}_{(r)}(\mathbf{p}) \cdot v_{(s)}(\mathbf{p})  = \tilde{v}_{(r)}(\mathbf{p}) \cdot u_{(s)}(\mathbf{p}) = 0
\end{align}
\end{subequations}
The indices $r,s \in \{1,2\}$ physically represent spin orientations of an electron or a positron. The functions $u_{(r)}(\mathbf{p})$ are physically associated with electron wave fields, whereas functions $v_{(r)}(\mathbf{p})$ are associated with the positron wave fields.
We have to consider in the later sections an almost static approximation to equations (21). In these approximations, we can derive that
\begin{subequations}
\begin{align}
& E(\mathbf{p}) = m +\left( ||\mathbf{p}||^2/2m \right) + O \left( ||\mathbf{p}||^4 \right),  \\
\notag\\
\notag\\
& u_{(1)} = 
\begin{bmatrix}
1\\
0\\
0\\
0
\end{bmatrix} +
\begin{bmatrix}
 (1/2)\left( ||\mathbf{p}||^2/2m \right)   \\
  0    \\
  -i(p_3/2m)[1-(1/2)\left( ||\mathbf{p}||/2m \right)^2 ] \\
  -i\left(\dfrac{p_1+ip_2}{2m}\right)[1-(1/2)\left( ||\mathbf{p}||/2m \right)^2 ]  \\
\end{bmatrix} + O \left( ||\mathbf{p}||^4 \right) \\
\notag\\
\notag\\
& u_{(2)} =
\begin{bmatrix}
0\\
1\\
0\\
0
\end{bmatrix} +
\begin{bmatrix}
 0  \\
 (1/2)\left( ||\mathbf{p}||^2/2m \right)   \\
  -i\left(\dfrac{p_1-ip_2}{2m}\right)[1-(1/2)\left( ||\mathbf{p}||/2m \right)^2 ] \\
  + i(p_3/2m)[1-(1/2)\left( ||\mathbf{p}||/2m \right)^2 ] \\
\end{bmatrix} + O \left( ||\mathbf{p}||^4 \right) \notag\\
\end{align}
\end{subequations}
Now, using Dirac matrices of equations (15) and equations (23), we deduce the almost static approximations 
\begin{subequations}
\begin{align}
\tilde{u}_{(r)}(\mathbf{p'}) \gamma^{a}  u_{(s)}(\mathbf{p}) & = 0 + O \left( ||\mathbf{p'}||^2 \right) +O \left( ||\mathbf{p}||^2 \right) \\
\tilde{u}_{(r)}(\mathbf{p'}) \gamma^4  u_{(s)}(\mathbf{p}) & = -i[\delta_{(rs)} + O \left( ||\mathbf{p'}||^2 \right) +O \left( ||\mathbf{p}||^2 \right)] 
\end{align}
\end{subequations}

A class of exact solutions of the Dirac Equations (18) is furnished by \cite{DasV,DasVI},
\begin{subequations}
\begin{align}
&\psi^{-}(\mathbf{n},t) = \int_{{\mathbb R}^3} d^3\mathbf{p}\,\,\,[m/E(\mathbf{p})]^{1/2}  
\left\{ \sum_{r=1}^{2}\alpha_{(r)}(\mathbf{p})u_{(r)}(\mathbf{p}) 
\left(\prod_{j=1}^{3} \xi_{n^j}(p_j) \right)  e^{-iE t} \right\} \\ 
&\psi^{+}(\mathbf{n},t) =  \int_{{\mathbb R}^3} d^3\mathbf{p}\,\,\,[m/E(\mathbf{p})]^{1/2}  
\left\{ \sum_{r=1}^{2}\beta^{\dagger}_{(r)}(\mathbf{p})v_{(r)}(\mathbf{p}) 
\left(\prod_{j=1}^{3} \overline{\xi_{n^j}(p_j)} \right)  e^{iE t} \right\} 
\\ & \psi(\mathbf{n},t) =  \psi^{-}(\mathbf{n},t) + \psi^{+}(\mathbf{n},t) \\
&\tilde{\psi}^{+}(\mathbf{n},t) = \int_{{\mathbb R}^3} d^3\mathbf{p}\,\,\,[m/E(\mathbf{p})]^{1/2}  
\left\{ \sum_{r=1}^{2}\alpha^{\dagger}_{(r)}(\mathbf{p})\tilde{u}_{(r)}(\mathbf{p}) 
 \left(\prod_{j=1}^{3} \overline{\xi_{n^j}(p_j)} \right)  e^{iE t} \right\} \\ 
&\tilde{\psi}^{-}(\mathbf{n},t) = \int_{{\mathbb R}^3} d^3\mathbf{p}\,\,\,[m/E(\mathbf{p})]^{1/2}  
\left\{ \sum_{r=1}^{2}\beta_{(r)}(\mathbf{p})\tilde{v}_{(r)}(\mathbf{p}) 
\left(\prod_{j=1}^{3} \xi_{n^j}(p_j) \right)  e^{-iE t} \right\} 
\\ & \tilde{\psi}(\mathbf{n},t) = \tilde{\psi}^{+}(\mathbf{n},t) +  \tilde{\psi}^{-}(\mathbf{n},t)
\end{align}
\end{subequations}

Now, we shall introduce the canonical or second quantization of the free electron-positron wave fields. We adopt a two-dimensional complex vector space called a Pre-Hilbert space. We postulate that the electron-positron field $\psi(\mathbf{n},t)$ of equation (25c) is a Pre-Hilbert space vector and that the Fourier coefficients $\alpha_{(r)}, \beta_{(r)}$ of equations (25a) and (25e) are linear operators acting on the  Pre-Hilbert space vectors. Morever, we assume that these linear operators satisfy anti-commutation rules \cite{DasV,DasVI},
\begin{subequations}
\begin{align}
[A,B]_+ :&= AB+BA=[B,A]_+ \\
[\alpha_{(r)}(\mathbf{p}), \alpha_{(s)}(\hat{\mathbf{p}})]_+ & = [\beta_{(r)}(\mathbf{p}), \beta_{(s)}(\hat{\mathbf{p}})]_+= \notag\\
[\alpha^{\dagger}_{(r)}(\mathbf{p}), \alpha^{\dagger}_{(s)}(\hat{\mathbf{p}})]_+ & = [\beta^{\dagger}_{(r)}(\mathbf{p}), \beta^{\dagger}_{(s)}(\hat{\mathbf{p}})]_+= 0\\
[\alpha_{(r)}(\mathbf{p}), \beta_{(s)}(\hat{\mathbf{p}})]_+ & = [\alpha^{\dagger}_{(r)}(\mathbf{p}), \beta^{\dagger}_{(s)}(\hat{\mathbf{p}})]_+= \notag\\
[\alpha_{(r)}(\mathbf{p}), \beta^{\dagger}_{(s)}(\hat{\mathbf{p}})]_+ & = [\alpha^{\dagger}_{(r)}(\mathbf{p}), \beta_{(s)}(\hat{\mathbf{p}})]_+= 0\\
[\alpha_{(r)}(\mathbf{p}), \alpha^{\dagger}_{(s)}(\hat{\mathbf{p}})] & = [\beta_{(r)}(\mathbf{p}), \beta^{\dagger}_{(s)}(\hat{\mathbf{p}})] = \delta_{(rs)} \delta^3 (\mathbf{p}-\hat{\mathbf{p}}) \mathbf{I}
\end{align}
\end{subequations}

There are three physical motivations for choosing Pre-Hilbert space operators $\alpha_{(r)}(\mathbf{p}),\alpha^{\dagger}_{(s)}(\hat{\mathbf{p}}),  \beta_{(r)}(\mathbf{p}), \beta^{\dagger}_{(s)}(\hat{\mathbf{p}})$ satisfying anti-commutation relations (26) in contrast to Hilbert space operators $a_{\mu}(\mathbf{k}), a^{\dagger}_{\nu}(\hat{\mathbf{k}})$ satisfying commutation relations (12): (I) A Photon field obeys Bose-Einstein statistics, whereas the electron-positron field obeys Fermi-Dirac statistics. (II) The total energy of the electron field or positron field is a positive definite operator.
(III) The total electric charge of the electron field is negative, whereas the total electric charge of the positron field is positive.

Now, we shall work out the anti-commutation relations between various electron-positron wave fields. We use equations (25) and (26) to deduce
\begin{subequations}
\begin{align}
& [\psi^{(-)}(\mathbf{n},t), \tilde{\psi}^{(-)}(\hat{\mathbf{n}},\hat{t})]_+  = [\psi^{(+)}(\mathbf{n},t), \tilde{\psi}^{(+)}(\hat{\mathbf{n}},\hat{t})]_+ = 0 \\
& [\psi(\mathbf{n},t), \psi(\hat{\mathbf{n}},\hat{t})]_+  = [\tilde{\psi}(\mathbf{n},t), \tilde{\psi}(\hat{\mathbf{n}},\hat{t})]_+ = 0 \\
& [\psi^{(-)}(\mathbf{n},t), \tilde{\psi}^{(+)}(\hat{\mathbf{n}},\hat{t})]_+ =  \int_{{\mathbb R}^3} d^3\mathbf{p}\,\,\,[m/E(\mathbf{p})] \notag\\ 
& \sum_{r=1}^{2}u_{(r)}(\mathbf{p})\tilde{u}_{(r)}(\mathbf{p})  \left(\prod_{j=1}^{3} \xi_{n^j}(p_j)
\overline{\xi_{\hat{n}^j}(p_j)} \right) e^{-iE(t-\hat{t})} \notag\\
& =: i S_{(+)}(\mathbf{n}, t ;\mathbf{\hat{n}},\hat {t};m) \mathbf{I} \\
& [\psi^{(+)}(\mathbf{n},t), \tilde{\psi}^{(-)}(\hat{\mathbf{n}},\hat{t})]_+  
= i S_{(-)}(\mathbf{n}, t ;\mathbf{\hat{n}},\hat {t};m) \mathbf{I} \\
& [\psi(\mathbf{n},t), \tilde{\psi}(\hat{\mathbf{n}},\hat{t})]_+  
= i S(\mathbf{n}, t ;\mathbf{\hat{n}},\hat {t};m) \mathbf{I} \\
& S(\mathbf{n}, t ;\mathbf{\hat{n}},\hat {t};m) :=S_{(+)}(\mathbf{n}, t ;\mathbf{\hat{n}},\hat {t};m) +S_{(-)}(\mathbf{n}, t ;\mathbf{\hat{n}},\hat {t};m) 
\end{align}
\end{subequations}
The various non-singular Green's functions $S_{(a)}(\mathbf{n}, t ;\mathbf{\hat{n}},\hat {t};m)$ will be elaborated in the Appendix.

\section{Interaction of various fields, $S^{\#}$-matrix, and M\o ller Scattering}

The relativistic Lagrangian of the interacting photon and Dirac fields is taken to be \cite{DasVII}
\begin{equation}
L_{(int)} (\mathbf{n}, t) := -ieN [ \tilde{\psi} (\hat{\mathbf{n}},\hat{t}) \gamma^{\mu} \psi(\mathbf{n},t) A_{\mu} (\mathbf{n},t)].
\end{equation}
Here, $|e| = \sqrt{4 \pi/137} < 1$, the electric charge of an electron is $-|e|<0$, whereas the elcetric charge of an positron $|e| >0$, and $N[ \cdots ]$ stands for normal ordering.

The scattering matrix in the discrete phase space and continuous time, denoted by $S^{\#}$-matrix, is defined by the operator-valued infinite series,
\begin{equation}
\begin{array}{l}
S^{\#} = I + \sum_{j=1}^{\infty}S^{\#}_{j} := I + \sum\limits_{j=1}^{\infty} \dfrac{(e)^j}{j!} \sum\limits_{\mathbf{n}_{(1)}=1}^{\infty (3)} \cdots \sum\limits_{\mathbf{n}_{(j)}=1}^{\infty (3)}  \int\limits_{\mathbb{R}} dt_{(1)}  \\
\cdots \int\limits_{\mathbb{R}} dt_{(j)}  T \left\{ N[\tilde{\psi} (\mathbf{n}_{(1)},t_{(1)}) \gamma^{\mu_1} \psi(\mathbf{n}_{(1)},t_{(1)}) A_{\mu_{(1)}} (\mathbf{n}_{(1)},t_{(1)})]  \right. \\
\cdots \left. N[\tilde{\psi} (\mathbf{n}_{(j)},t_{(j)}) \gamma^{\mu_j} \psi(\mathbf{n}_{(1)},t_{(j)}) A_{\mu_{(j)}} (\mathbf{n}_{(j)},t_{(j)})] \right\}
\end{array}
\end{equation}

Here, $T$ denotes Wick's time ordering operation. We distinguish the scattering matrix by the notation $S^{\#}$-matrix from the usual notation of S-matrix in continuous space-time because the physics in the discrete phase space and continuous time is different from the physics in the space-time continuum.

We provide Feynman rules to evaluate succinctly each term of the $S^{\#}$-matrix series in (29)  in Tables 1 and 2.

\begin{table}[h]
\caption{Feynman Rules} 
\centering 
\begin{tabular}{c c } 
\hline\hline 
Description & Factor in  $S^{\#}$-matrix  \\ [0.5ex] 
\hline 
Photon propagator & $-i\eta_{\mu \nu}D_{(F+)}(\mathbf{n}, t ;\mathbf{\hat{n}},\hat {t})$  \\ 
Electron-positron propagator & $iS_{(F+)}(\mathbf{n}, t ;\mathbf{\hat{n}},\hat {t};m)$  \\
Electron-photon vertex & $\gamma^{\mu}$ \\
Incoming or outgoing external photon lines: & $A^{(-)}_{\mu}(\mathbf{n}, t)$ or $A^{(+)}_{\mu}(\mathbf{n}, t)$   \\
Incoming or outgoing external electron lines: & $\psi^{(-)}(\mathbf{n}, t)$ or $\tilde{\psi}^{(+)}(\mathbf{n}, t)$ \\ 
Incoming or outgoing external positron lines: & $\tilde{\psi}^{(-)}(\mathbf{n}, t)$ or $\psi^{(+)}(\mathbf{n}, t)$ \\ [1ex] 
\hline 
\end{tabular}
\label{table:Table-I} 
\end{table}

\begin{table}
\begin{center}
\includegraphics[scale=0.45]{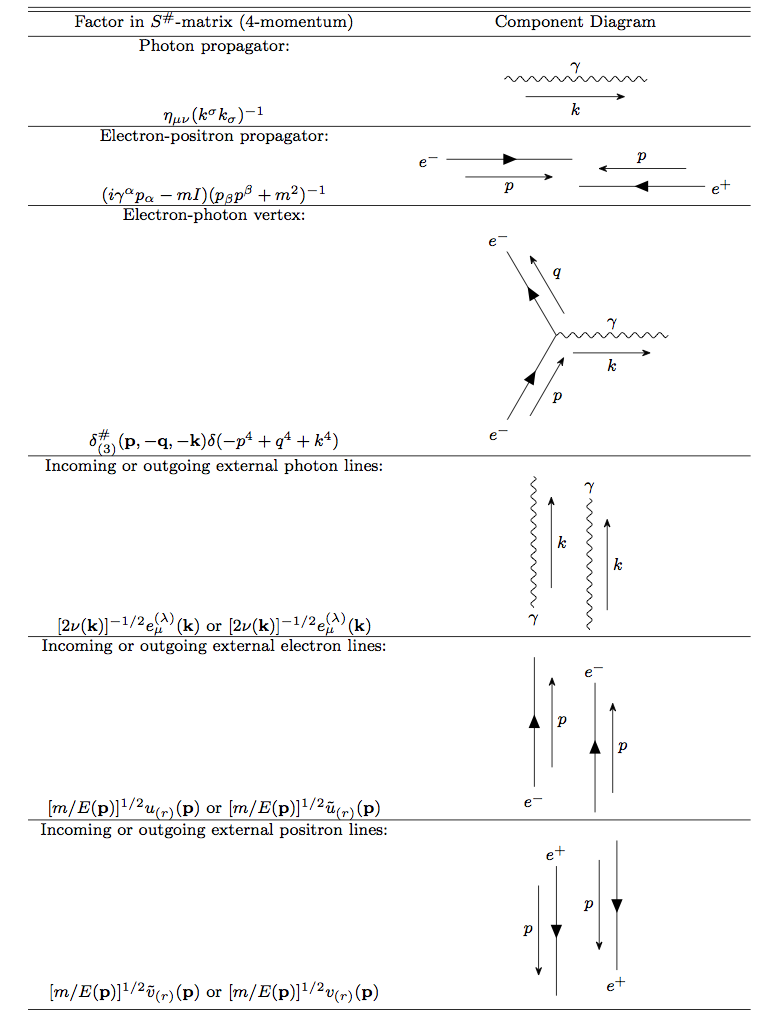}
\end{center}
\caption{Feynman Rules Continued}
\end{table}

The actual construction of the $S^{\#}_{(j)}$ element from Tables 1 and 2 is incomplete until we determine the appropriate numerical factor $c^{\#}_{(j)}$  to be multiplied to the $j$-th term, It turns out that the correct factor is 
\begin{equation}
c^{\#}_{(j)} = (-1)^l \delta_p (2\pi)^{j-(P_i+E_i)} (e)^{j}
\end{equation}
Here, $j$ is the number of vertices,  $(-1)^l$ is associated with each loop, and $P_i+E_i$ is the sum of internal photon and electron propagators.

Moreover, the vertex factor in 4-momentum space is defined by
\begin{subequations}
\begin{align}
& \delta^{\#}_{(3)}(\mathbf{p},-\mathbf{q},-\mathbf{k}) := \sum_{\mathbf{n}=0}^{\infty (3)} \left[\prod_{j=1}^{3} \xi_{n^j}(p_j) \overline{\xi_{n^j}(q_j)} \overline{\xi_{n^j}(k_j)} \right]\\
& \delta^{\#}_{(3)}(\mathbf{p},\mathbf{q},\mathbf{k}) := \sum_{\mathbf{n}=0}^{\infty (3)} \left[\prod_{j=1}^{3} \xi_{n^j}(p_j) \xi_{n^j}(q_j) \xi_{n^j}(k_j) \right]
\end{align}
\end{subequations}
The function $\delta^{\#}_{(3)}(\mathbf{p},-\mathbf{q},-\mathbf{k})$ above is different from the standard delta function factor $(2 \pi)^3 \delta^{3}(\mathbf{p},-\mathbf{q},-\mathbf{k})$ found in the usual theory in space-time continuum. Thus, in the discrete phase space theory, exact conservation of the total 3-momentum in a vertex is violated. In the quantum field theory of interactions over any discrete space, exact conservation of momentum in a vertex is lost \cite{DasI,Maradudin}.

If we denote the initial and final state vectors by $|i \rangle$ and $|f \rangle$ of a physical process, then from (29) we can define
\begin{equation}
\langle f | S^{\#} - I | i \rangle =: i (2 \pi) \delta (E_{(f)} - E_{(i)}) \langle f | M^{\#} | i \rangle
\end{equation}
Here, $E_{(i)}$ stands for the total initial energy, whereas $E_{(f)}$ stands for the total final energy. The equation (32) implies the exact conservation of the total energy in the interaction. The transition probability from the initial state $|i \rangle$ to the final state $|f \rangle$ per unit time is provided by 
\begin{equation}
w_{(fi)}=: i (2 \pi)^2 \delta (E_{(f)} - E_{(i)}) | \langle f | M^{\#} | i \rangle |^2
\end{equation}
("Fermi's golden rule").

As an illustration of computing the second order element $\langle f | S^{\#}_{(2)} - I | i \rangle$ in (29), we consider the case of the electron-electron scattering or M\o ller scattering. The corresponding component of the Feynman diagram in 4-momentum space is exhibited in Figure 1.

\begin{figure}
\begin{center}
\includegraphics[scale=0.50]{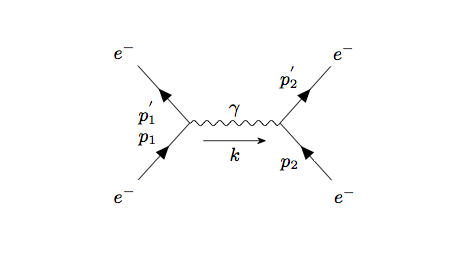}
\end{center}
\caption{The second order Feynman diagram for M\o ller scattering}
\end{figure}

Using the 4-momentum version of Table 1, 2, and equation (30), we obtain the required second order $S^{\#}$-matrix element as,
\begin{subequations}
\begin{align}
&\langle p^{'}_2 , p^{'}_1 | S^{\#}_{(2)} | p_1 , p_2  \rangle = [c^{\#}_2] \left[ \dfrac{m^2}{\sqrt{E^{'}_1 E^{'}_2 E_1 E_2}} \right] \notag\\ 
&\int_{{\mathbb R}^4} \left\{ [\tilde{u}(\mathbf{p^{'}_1}) \gamma^{\mu} u(\mathbf{p_1})] \delta^{\#}_{(3)}(\mathbf{p_1},-\mathbf{p^{'}_1},-\mathbf{k}) \delta(p^4_1 - p^{'4}_1 - k^4) \right. \notag\\
& \left[ \dfrac{\eta_{\mu \nu}}{\mathbf{k} \cdot \mathbf{k} -(k^4 )^2} \right] [\tilde{u}(\mathbf{p^{'}_2}) \gamma^{\nu} u(\mathbf{p_2})] \delta^{\#}_{(3)}(\mathbf{p_2},-\mathbf{p^{'}_2},+\mathbf{k}) \notag\\
& \left. \delta(p^4_2 - p^{'4}_2 + k^4) \right\} d^3\mathbf{k} dk^4 \\ 
&\langle p^{'}_2 , p^{'}_1 | S^{\#}_{(2)} | p_1 , p_2  \rangle = \left[ \dfrac{-i(2\pi)e^2 m^2}{\sqrt{E^{'}_1 E^{'}_2 E_1 E_2}} \right] \delta(E^{'}_2+E^{'}_1-E_2-E_1)  \notag\\ 
&\int_{{\mathbb R}^3} \left\{ [\tilde{u}(\mathbf{p^{'}_1}) \gamma^{\mu} u(\mathbf{p_1})] \delta^{\#}_{(3)}(\mathbf{p_1},-\mathbf{p^{'}_1},-\mathbf{k}) \left[ \dfrac{\eta_{\mu \nu}}{\mathbf{k} \cdot \mathbf{k} -(E^{'}_1-E_1 )^2} \right] \right. \notag\\
&  \left. [\tilde{u}(\mathbf{p^{'}_2}) \gamma^{\nu} u(\mathbf{p_2})] \delta^{\#}_{(3)}(\mathbf{p_2},-\mathbf{p^{'}_2},+\mathbf{k})  \right\} d^3\mathbf{k} 
\end{align}
\end{subequations} 

To compare and contrast the above equations (34) with the usual second order S-matrix elements for M\o ller scattering over space-time continuum, we furnish from \cite{Jauch}
\begin{subequations}
\begin{align}
&\langle p^{'}_2 , p^{'}_1 | S_{(2)} | p_1 , p_2  \rangle = [c_{(2)}] \left[ \dfrac{m^2}{\sqrt{E^{'}_1 E^{'}_2 E_1 E_2}} \right] \notag\\ 
&\int_{{\mathbb R}^4} \left\{ [\tilde{u}(\mathbf{p^{'}_1}) \gamma^{\mu} u(\mathbf{p_1})] \delta^{\#}_{(3)}(\mathbf{p_1},-\mathbf{p^{'}_1},-\mathbf{k}) \delta(p^4_1 - p^{'4}_1 - k^4) \right. \notag\\
& \left[ \dfrac{\eta_{\mu \nu}}{\mathbf{k} \cdot \mathbf{k} -(k^4 )^2} \right] [\tilde{u}(\mathbf{p^{'}_2}) \gamma^{\nu} u(\mathbf{p_2})] \delta^{\#}_{(3)}(\mathbf{p_2},-\mathbf{p^{'}_2},+\mathbf{k}) \notag\\
& \left. \delta(p^4_2 - p^{'4}_2 + k^4) \right\} d^3\mathbf{k} dk^4 \\ 
&\langle p^{'}_2 , p^{'}_1 | S_{(2)}  | p_1 , p_2  \rangle =  \dfrac{-(ie^2 m^2)/(4\pi^2)}{\sqrt{E^{'}_1 E^{'}_2 E_1 E_2}}\delta(E^{'}_2+E^{'}_1-E_2-E_1)  \notag\\ 
&\int_{{\mathbb R}^3} \left\{ [\tilde{u}(\mathbf{p^{'}_1}) \gamma^{\mu} u(\mathbf{p_1})] \delta^{\#}_{(3)}(\mathbf{p_1},-\mathbf{p^{'}_1},-\mathbf{k}) \dfrac{\eta_{\mu \nu}}{\mathbf{k} \cdot \mathbf{k} -(E^{'}_1-E_1 )^2}  \right. \notag\\
&  \left. [\tilde{u}(\mathbf{p^{'}_2}) \gamma^{\nu} u(\mathbf{p_2})] \delta^{\#}_{(3)}(\mathbf{p_2},-\mathbf{p^{'}_2},+\mathbf{k})  \right\} d^3\mathbf{k}  
\end{align}
\end{subequations}
The last equation (35) can be further reduced to the following algebraic form \cite{Jauch},
\begin{subequations}
\begin{align}
&\langle p^{'}_2 , p^{'}_1 | S_{(2)}  | p_1 , p_2  \rangle =  \dfrac{-(ie^2 m^2)/(4\pi^2)}{\sqrt{E^{'}_1 E^{'}_2 E_1 E_2}} \delta(p^{'}_2+p^{'}_1-p_2-p_1)  \notag\\ 
&  [\tilde{u}(\mathbf{p^{'}_1}) \gamma^{\mu} u(\mathbf{p_1})]  \dfrac{\eta_{\mu \nu}}{\eta^{\alpha \beta} (p^{'}_{1\alpha}-p_{1\alpha})(p^{'}_{1\beta}-p_{1\beta})}  [\tilde{u}(\mathbf{p^{'}_2}) \gamma^{\nu} u(\mathbf{p_2})] 
\end{align}
\end{subequations}
Such a drastic reduction for equation (34) in the discrete case is not possible at the current time. Let us consider the usual M\o ller scattering formula (35b) in the continuous case. Using the relation $\delta^3(\mathbf{p}-\mathbf{p}^{'}-\mathbf{k}) = (2 \pi)^{-3}  \int_{{\mathbb R}^3} e^{i\mathbf{p} \cdot \mathbf{x}} d^3 \mathbf{x}$, (35b) becomes
\begin{equation}
\begin{array}{c}
\langle p^{'}_2 , p^{'}_1 | S_{(2)}  | p_1 , p_2  \rangle = 
\dfrac{-(ie^2 m^2)/(4\pi^2)}{\sqrt{E^{'}_1 E^{'}_2 E_1 E_2}} \delta(E^{'}_2+E^{'}_1-E_2-E_1)  \\ 
\int_{{\mathbb R}^3} \left\{ [\tilde{u}(\mathbf{p^{'}_1}) \gamma^{\mu} u(\mathbf{p_1})]  \int_{{\mathbb R}^3} e^{i(\mathbf{p_1}-\mathbf{p^{'}_1}-\mathbf{k}) \cdot \mathbf{x_1}} \dfrac{d^3 \mathbf{x_1}}{(2 \pi)^{3}} \right. \left[ \dfrac{\eta_{\mu \nu}}{\mathbf{k} \cdot \mathbf{k} -(E^{'}_1-E_1 )^2} \right] \\
  \left. [\tilde{u}(\mathbf{p'_2}) \gamma^{\nu} u(\mathbf{p_2})] \int_{{\mathbb R}^3} e^{i(\mathbf{p_2}-\mathbf{p^{'}_2}+\mathbf{k}) \cdot \mathbf{x_2}} \dfrac{d^3 \mathbf{x_2}}{(2 \pi)^{3}}  \right\} d^3\mathbf{k}  
\end{array}
\end{equation}
Now, we assume slow motions of the two external electrons and consequent equations (23) and (24). We also assume conservation of electron spin implying $r_1^{'} =r_1$ and $r_2^{'} =r_2$. Then equation (37) reduces to 
\begin{equation}
\begin{array}{c}
\langle p^{'}_2 , p^{'}_1 | S_{(2)}  | p_1 , p_2  \rangle = 
\dfrac{-i e^2}{(2\pi)^{5}} \delta\left(\dfrac{||\mathbf{p'_1}||^2}{2m}+\dfrac{||\mathbf{p'_2}||^2}{2m}-\dfrac{||\mathbf{p_1}||^2}{2m}-\dfrac{||\mathbf{p_2}||^2}{2m} \right)  \\ 
\int_{{\mathbb R}^3} \int_{{\mathbb R}^3} e^{i(\mathbf{p_1}-\mathbf{p^{'}_1}) \cdot \mathbf{x_1}}  e^{i(\mathbf{p_2}-\mathbf{p^{'}_2}) \cdot \mathbf{x_2}} 
\int _{{\mathbb R}^3}  \dfrac{e^{-i(\mathbf{x_1}-\mathbf{x_2})\cdot \mathbf{k}}} {(2 \pi)^3[\mathbf{k} \cdot \mathbf{k}]} d^3\mathbf{k} d^3 \mathbf{x_1} d^3 \mathbf{x_2} \\
+ (\textbf{Higher order terms})
\end{array}
\end{equation}
We can read off from the above equations the Green's function $G(\mathbf{x_1}-\mathbf{x_2})$ of the usual potential equation as,
\begin{subequations}
\begin{align}
& G(\mathbf{x_1}-\mathbf{x_2}) = \int_{{\mathbb R}^3}  \dfrac{e^{-i(\mathbf{x_1}-\mathbf{x_2})\cdot \mathbf{k}}} {(2 \pi)^3[\mathbf{k} \cdot \mathbf{k}]} d^3\mathbf{k} \\ 
& G(\mathbf{x_1}-\mathbf{x_2}) =\dfrac{1}{(4\pi) || \mathbf{x_1}-\mathbf{x_2} ||}  \\ 
& \delta^{ab} \dfrac{\partial^2}{\partial x^a_{1} \partial x^b_{2}} G(\mathbf{x_1}-\mathbf{x_2}) =-\delta^{3}(\mathbf{x_1}-\mathbf{x_2})  \\
& \lim_{\mathbf{x_1} \rightarrow \mathbf{x_2}} G(\mathbf{x_1}-\mathbf{x_2}) \rightarrow \infty
\end{align}
\end{subequations}
We obtain the usual singular Coulomb potential between two electrons as \cite{Peskin},
\begin{equation}
V(\mathbf{x_1},\mathbf{x_2}) = e^2 G(\mathbf{x_1}-\mathbf{x_2}) =\dfrac{e^2}{(4\pi) || \mathbf{x_1}-\mathbf{x_2} ||} 
\end{equation}

Suppose that one of the charged particles has unit electric charge and is situated at the origin $\mathbf{x_2}=(0,0,0)$. Let the other particle have electric charge of $-1$ and be situated at $\mathbf{x_1}=(x^1,0,0)$. the corresponding Coulomb potential is given by,
\begin{subequations}
\begin{align}
& W(x^1,0,0) :=V(\mathbf{x_1},\mathbf{0}) = -\dfrac{1}{(4\pi) |x^1|} < 0 \\ 
& \lim_{x^1 \rightarrow \infty} W(x^1,0,0) = 0
\end{align}
\end{subequations}

Now we shall examine the matrix element $\langle p^{'}_2 , p^{'}_1 | S^{\#}_{(2)} | p_1 , p_2  \rangle $ for M\o ller scattering in discrete phase space. Using equations (31) and (34), we obtain 
\begin{equation}
\begin{array}{c}
\langle p^{'}_2 , p^{'}_1 | S^{\#}_{(2)} | p_1 , p_2  \rangle  \dfrac{-(ie^2 2\pi m^2)}{\sqrt{E^{'}_1 E^{'}_2 E_1 E_2}}  \delta(E^{'}_2+E^{'}_1-E_2-E_1)   \\ 
\int_{{\mathbb R}^3} \left\{ [\tilde{u}(\mathbf{p^{'}_1}) \gamma^{\mu} u(\mathbf{p_1})] \sum_{\mathbf{n}=0}^{\infty (3)} \left[\prod_{a=1}^{3} \xi_{n^a}(p_{1a}) \overline{\xi_{n^a}(p'_{1a})} \overline{\xi_{n^a}(k_a)} \right] \right.\\
\dfrac{\eta_{\mu \nu}}{\mathbf{k} \cdot \mathbf{k} -(E^{'}_1-E_1 )^2} [\tilde{u}(\mathbf{p^{'}_2}) \gamma^{\nu} u(\mathbf{p_2})] \\
\left. \sum_{\mathbf{\hat{n}}=0}^{\infty (3)} \left[\prod_{b=1}^{3} \xi_{\hat{n}^b}(p_{1b}) \overline{\xi_{\hat{n}^b}(p'_{1b})} \overline{\xi_{\hat{n}^b}(k_a)} \right] \right\}d^3\mathbf{k}
\end{array}
\end{equation}

Now we assume slow momenta for two external electrons and consequent equations (23) and (24). Then (42) reduces to 
\begin{equation}
\begin{array}{c}
\langle p^{'}_2 , p^{'}_1 | S^{\#}_{(2)} | p_1 , p_2  \rangle = 
-i e^2 \delta\left(\dfrac{||\mathbf{p'_1}||^2}{2m}+\dfrac{||\mathbf{p'_2}||^2}{2m}-\dfrac{||\mathbf{p_1}||^2}{2m}-\dfrac{||\mathbf{p_2}||^2}{2m} \right)  \\ 
\sum_{\mathbf{n}=0}^{\infty (3)} \sum_{\mathbf{\hat{n}}=0}^{\infty (3)}  \left[\prod_{a=1}^{3} \xi_{n^a}(p_{1a}) \overline{\xi_{n^a}(p'_{1a})}  \right] \left[\prod_{b=1}^{3} \xi_{\hat{n}^b}(p_{1b}) \overline{\xi_{\hat{n}^b}(p'_{1b})}\right]\\
\int_{{\mathbb R}^3} \left[\prod_{j=1}^{3} \overline{\xi_{n^j}(k_{j})} \xi_{\hat{n}^j}(k_{j})(\mathbf{k} \cdot \mathbf{k})^{-1} \right] d^3\mathbf{k} \\
+ (\textbf{Higher order terms})
\end{array}
\end{equation}
Comparing (43) to (38), we decduce that the relevant Green's function must be \cite{DasIX}
\begin{subequations}
\begin{align}
& G^{\#}(\mathbf{n}; \mathbf{\hat{n}}) = \int_{{\mathbb R}^3} \left[\prod_{j=1}^{3} \overline{\xi_{n^j}(k_{j})} \xi_{\hat{n}^j}(k_{j})(\mathbf{k} \cdot \mathbf{k})^{-1} \right] d^3\mathbf{k} \\ 
& \delta^{ab} \Delta^{\#}_a \Delta^{\#}_b G^{\#}(\mathbf{n}, \mathbf{\hat{n}})  =-\left[ \prod_{j=1}^{3} \delta_{n^j \hat{n}^j} \right] \\ 
& G^{\#}(0; 0) = 2
\end{align}
\end{subequations}
Comparing (44) with (39), we obtain the new non-singular Coulomb potential as in \cite{DasIX},
\begin{equation}
V^{\#}(\mathbf{n}, \mathbf{\hat{n}}) = e^2 G^{\#}(\mathbf{n}, \mathbf{\hat{n}})
\end{equation}
Suppose that one of the charged particles has unit electric charge and is situated at the discrete origin $\mathbf{\hat{n}}=(0,0,0)$. Let the other particle have electric charge of $-1$ and be situated at the discrete point $\mathbf{n}=(n^1,0,0)$. the corresponding new Coulomb potential is furnished by,
\begin{subequations}
\begin{align}
& W^{\#}(2n^1,0,0) := -G^{\#}(2n^1,0,0; 0,0,0) \notag\\
& = -\left[ \dfrac{2^{n^1 +1} (n^1 !)}{(2n^1+1)!\sqrt{(2n^1)!}} \right] < 0,  \\ 
& W^{\#}(2n^1+1,0,0)  = 0 \\
& \left[ \dfrac{W^{\#}(2n^1+2,0,0)}{W^{\#}(2n^1,0,0)} \right] 
= \left[ \dfrac{(2n^1)^2+6n^1+2}{(2n^1)^2+12n^1+9}  \right]^{1/2} < 1 \\
& \lim_{n^1 \rightarrow \infty}  W^{\#}(2n^1,0,0) = 0, \\
& W^{\#}(0,0,0) = -2
\end{align}
\end{subequations}

Thus we can conclude from (46) that the sequence $ \left\{ W^{\#}(2n^1,0,0) \right\}_0^{\infty}$ is a monotone increasing sequence of negative numbers. Comparing and contrasting Coulomb potentials $W(x^1,0,0)$ of ordinary space and $W^{\#}(n^1,0,0)$ of discrete phase space from equations (41) and (46) respectively, we exhibit Figure 2 below.

\begin{figure}
\setcounter{figure}{1}
\begin{flushleft}
\includegraphics[scale=0.70]{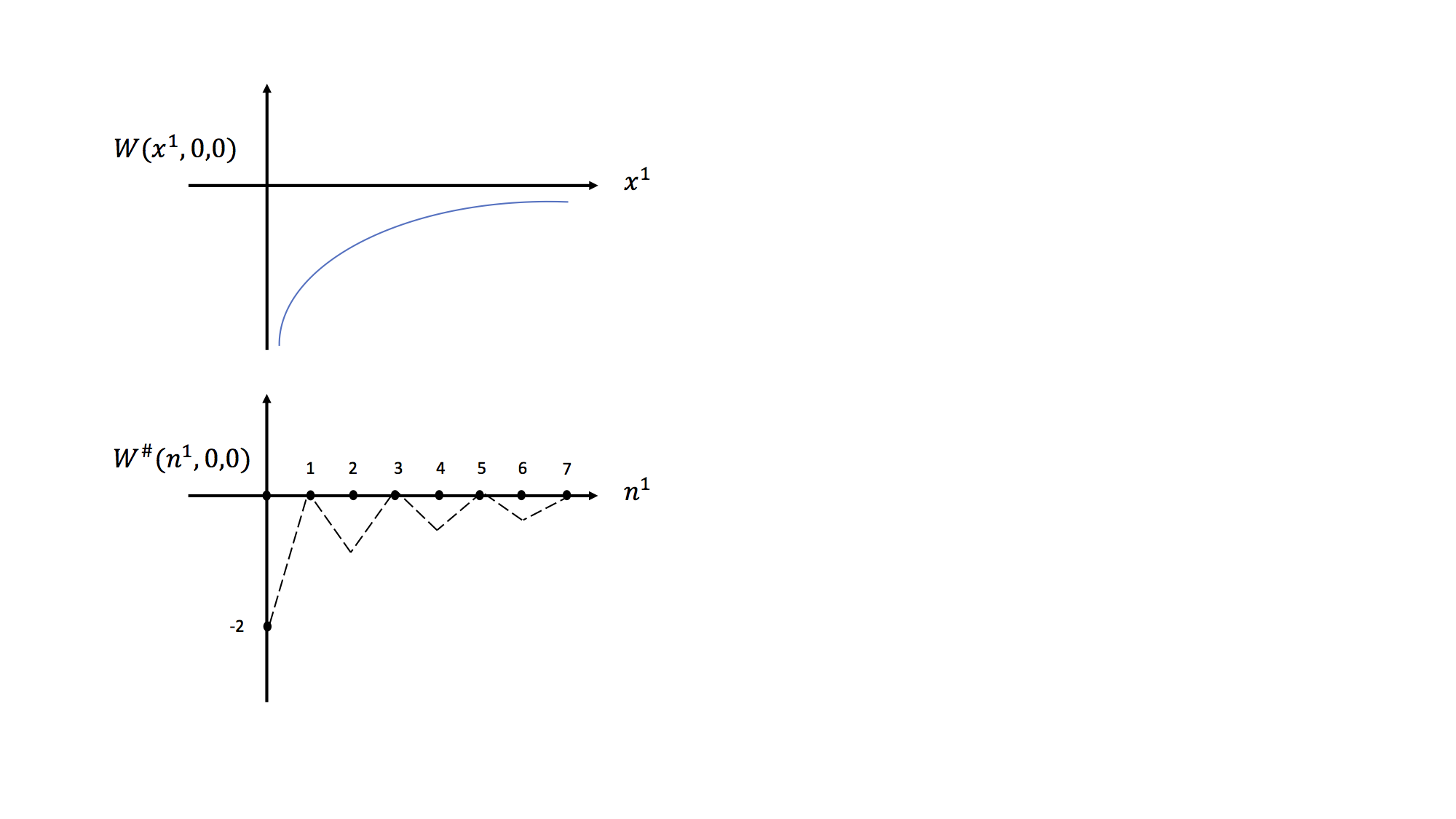}
\end{flushleft}
\caption{A graph of the usual singular potential Coulomb potential well $W(x^1,0,0)$ versus the new non-singular Coulomb potential well $W^{\#}(n^1,0,0)$ in discrete phase space.}
\end{figure}

\section{Concluding Remarks}

Let us physically analyze the usual Coulomb potential $W(x^1,0,0)$  in three dimensional physical space  
${\mathbb R}^3$ represented by the top graph of Figure 2. On Dirac particle of electric charge $+1$ is residing at the origin $(0,0,0)$. Another Dirac particle of electric charge $-1$ is residing at $(x^1,0,0)$ falling off in the infinite hole of the singular potential well characterized by  $W(x^1,0,0) = -\dfrac{1}{(4\pi) |x^1|}$. On the other hand, the bottom graph of Figure 2, the Dirac particle of electric charge $+1$ is residing at $(0,0,0)$ of discrete phase space. The other Dirac particle of electric charge $-1$ at $(n^1,0,0)$ falls off along the non-singular potential well at a finite depth characterized by the non-singular Coulomb potential $W^{\#}(n^1,0,0)$. However, this particle falls off along a "zig-zag" trajectory. Such a trajectory was called a \textit{Zitterbewegung} a long time ago.

\section*{Appendix: Discrete phase space, continuous time, and non-singular Green's functions for free relativistic field equations}

The relativistic partial difference-differential equation for a real scalar field (or Klein-Gordon field) is provided by 
\begin{equation}
\delta^{jl} \Delta_j^{\#}\Delta_l^{ \#} \phi(\mathbf{n},t) -(\partial_t)^2 \phi(\mathbf{n},t) -\mu^2 \phi(\mathbf{n},t) = 0
\end{equation}
Here, $\mu >0$ is the mass parameter.

The associated Green's function are given by \cite{DasV,DasVI},
\begin{equation}
\begin{array}{c}
\Delta_{(a)}(\mathbf{n},t; \mathbf{\hat{n}},\hat{t};\mu) = (2\pi)^{-1} \int_{{\mathbb R}^3} \left\{
 \left[\prod_{j=1}^{3} \xi_{n^j}(k_j) \overline{\xi_{\hat{n}^j}(k_j)}  \right] \right. \\
\left. \int_{C_{(a)}} \left[ \left( \eta^{\alpha \beta} k_{\alpha} k_{\beta} +\mu^2 \right)^{-1}
\mathbf{exp}(ik_4(t-\hat{t})]dk^4 \right]  \right\} d^3\mathbf{k} 
\end{array}
\end{equation}
(Note that in the signature +2 convention, $k^4 = -k_4$.) The Green's functions above involve nine contours in the complex $k^4$-plane \cite{Jauch}. These contours are exhibited explicitly in Figure 3 with $w = w(\mathbf{k}) := +\sqrt{\mathbf{k} \cdot \mathbf{k} +\mu^2} > 0$.

\begin{figure}
\begin{flushleft}
\includegraphics[scale=0.35]{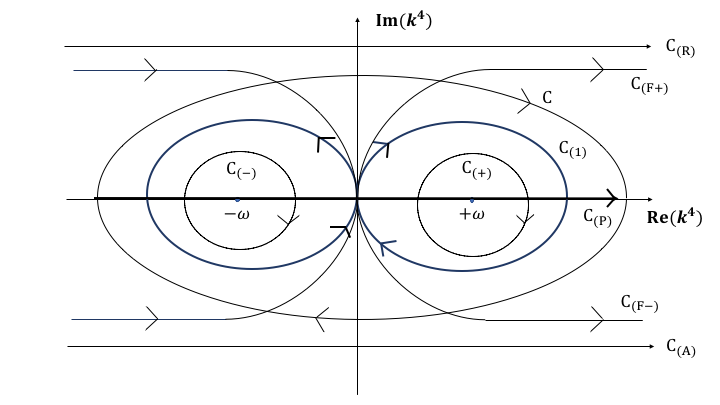}
\end{flushleft}
\caption{Various contours $C_{(a)}$ in the complex $k^4$ plane.}
\end{figure}

It can be verified that 
\begin{equation}
\begin{array}{c}
\delta^{jl} \Delta_j^{\#}\Delta_l^{ \#} \Delta_{(a)}(\mathbf{n},t; \mathbf{\hat{n}},\hat{t}) -(\partial_t)^2 \phi(\mathbf{n},t) -\mu^2 \Delta_{(a)}(\mathbf{n},t; \mathbf{\hat{n}},\hat{t})=0 \\
 \textbf{for contours} \,\,\,  C_{(a)}=C, C_{(+)},C_{(-)},C_{(1)} \\
 \\
 \\
= -\delta_{n^1 \hat{n}^1} \delta_{n^2 \hat{n}^2} \delta_{n^3 \hat{n}^3}\delta(t-\hat{t})\,\,\, \notag\\
\textbf{for contours} \,\,\, C_{(a)}=C_{(R)}, C_{(A)},C_{(P)},C_{(F+)},C_{(F-)}  \\
\end{array}
\end{equation}

The Green's functions $D_{(a)}(\mathbf{n},t;\hat{\mathbf{n}},\hat{t})$ in equations (14) for a photon field are defined by \cite{DasV,DasVI},
\begin{equation}
\begin{array}{c}
D_{(a)}(\mathbf{n},t;\hat{\mathbf{n}},\hat{t}):=\Delta_{(a)}(\mathbf{n},t;\hat{\mathbf{n}},\hat{t}:0) =
\dfrac{1}{2\pi} \int_{{\mathbb R}^3} \left\{
 \left[\prod_{j=1}^{3} \xi_{n^j}(k_j) \overline{\xi_{\hat{n}^j}(k_j)}  \right] \right. \\
\left. \int_{C_{(a)}} \left[ \left( \eta^{\alpha \beta} k_{\alpha} k_{\beta} \right)^{-1}
\mathbf{exp}(ik_4(t-\hat{t})]dk^4 \right] \right\} d^3\mathbf{k} 
\end{array}
\end{equation}
Here, in Figure 3, we have to replace $w=w(\mathbf{k})$ by $\nu=\nu(\mathbf{k}):=\sqrt{\mathbf{k} \cdot \mathbf{k}}$.
By carrying out some closed contour integrals over the complex $k^4$-plane in (49), one arrives at the following three-dimensional representations,
\begin{subequations}
\begin{align}
& D_{(+)}(\mathbf{n},t;\hat{\mathbf{n}},\hat{t}) = i \int_{{\mathbb R}^3} 
\left[\prod_{j=1}^{3} \xi_{n^j}(k_j) \overline{\xi_{\hat{n}^j}(k_j)}  \right] \dfrac{e^{-i\nu(t-\hat{t})}}{\nu(\mathbf{k})}d^3\mathbf{k} \\
& D_{(-)}(\mathbf{n},t;\hat{\mathbf{n}},\hat{t}) = -i \int_{{\mathbb R}^3} 
\left[\prod_{j=1}^{3} \overline{\xi_{n^j}(k_j)} \xi_{\hat{n}^j}(k_j)  \right] \dfrac{e^{+i\nu(t-\hat{t})}}{\nu(\mathbf{k})}d^3\mathbf{k} \\
& D(\mathbf{n},t;\hat{\mathbf{n}},\hat{t}) = D_{(+)}(\mathbf{n},t;\hat{\mathbf{n}},\hat{t}) + D_{(-)}(\mathbf{n},t;\hat{\mathbf{n}},\hat{t})\notag\\
& =\int_{{\mathbb R}^3} 
\left[\prod_{j=1}^{3} \xi_{n^j}(k_j) \overline{\xi_{\hat{n}^j}(k_j)}  \right] \dfrac{\sin(\nu(t-\hat{t}))}{\nu}d^3\mathbf{k} \\
& D(\mathbf{n},t;\hat{\mathbf{n}},\hat{t})|_{t=\hat{t}} =0 \\
& D_{(1)}(\mathbf{n},t;\hat{\mathbf{n}},\hat{t}) =-i[ D_{(+)}(\mathbf{n},t;\hat{\mathbf{n}},\hat{t}) - D_{(-)}(\mathbf{n},t;\hat{\mathbf{n}},\hat{t})]\notag\\
& =\int_{{\mathbb R}^3} 
\left[\prod_{j=1}^{3} \xi_{n^j}(k_j) \overline{\xi_{\hat{n}^j}(k_j)}  \right] \dfrac{\cos(\nu(t-\hat{t}))}{\nu}d^3\mathbf{k} 
\end{align}
\end{subequations}
Now, Green's functions for Dirac field equations (18a) are provided by \cite{DasV,DasVI},
\begin{subequations}
\begin{align}
& S_{(a)}(\mathbf{n},t;\hat{\mathbf{n}},\hat{t};m) = (\gamma^b \Delta^{\#}_{b}  + \gamma^{4} \partial_t - mI) \Delta_{(a)}(\mathbf{n},t;\hat{\mathbf{n}},\hat{t};m)\\
& S_{(a)AB}(\mathbf{n},t;\hat{\mathbf{n}},\hat{t};m) = (\gamma^b_{AB} \Delta^{\#}_{b}  + \gamma^{4}_{AB} \partial_t - m\delta_{AB})
\Delta_{(a)}(\mathbf{n},t;\hat{\mathbf{n}},\hat{t};m)\\
& S_{(a)}(\mathbf{n},t;\hat{\mathbf{n}},\hat{t}) =(2\pi)^{-1}\int_{{\mathbb R}^3} \left\{ 
\left[\prod_{j=1}^{3} \xi_{n^j}(p_j) \overline{\xi_{\hat{n}^j}(p_j)}  \right] \right. \notag \\
& \left. \int_{C{(a)}} \dfrac{(i\gamma^{\mu}p_{\mu} -mI)e^{-ip^4(t-\hat{t})}}{\eta^{\alpha \beta}p_{\alpha}p_{\beta}+m^2}dp^{4} \right\} d^3\mathbf{p}
\end{align}
\end{subequations}

Here, $A,B \in \{1,2,3,4\}$ are bispinor indices. The $ 4\times 4$ matrix Green's functions (51) satisfy,
\begin{equation}
\begin{array}{c}
(\gamma^b \Delta^{\#}_{b}  + \gamma^{4} \partial_t + mI) S_{(a)}(\mathbf{n},t;\hat{\mathbf{n}},\hat{t};m) = [0]_{4\times 4}\\
 \textbf{for contours} \,\,\, C, C_{(+)},C_{(-)},C_{(1)} \\
 \\
 \\
= -\delta_{n^1 \hat{n}^1} \delta_{n^2 \hat{n}^2} \delta_{n^3 \hat{n}^3} \delta(t-\hat{t}) [1]_{4\times 4}\,\,\, \notag\\ 
\textbf{for contours} \,\,\, C_{(R)}, C_{(A)},C_{(P)},C_{(F+)},C_{(F-)}  
\end{array}
\end{equation}




\end{document}